\documentclass[12pt]{iopart}

\usepackage{amssymb}
\usepackage{graphicx}
\usepackage{iopams}

\newcommand{\bra}[1]{\ensuremath{\left<#1\right|}}
\newcommand{\ket}[1]{\ensuremath{\left|#1\right>}}
\newcommand{\oost}{\frac{1}{\sqrt{2}}}

\def\id{{\rm 1\kern-.22em l}}

\begin{document}

\title[Optimal witnesses for three-qubit entanglement from 
       GHZ symmetry]
       {Optimal witnesses for three-qubit entanglement from 
       Greenberger-Horne-Zeilinger symmetry}

\author{Christopher Eltschka}
\address{Institut f\"ur Theoretische Physik, 
         Universit\"at Regensburg, D-93040 Regensburg, Germany}
\author{Jens Siewert}
\address{Departamento de Qu\'{\i}mica F\'{\i}sica, 
         Universidad del Pa\'{\i}s Vasco UPV/EHU,
         48080 Bilbao, Spain}
\address{IKERBASQUE, Basque Foundation for Science, 48011 Bilbao, Spain}

\begin{abstract}            
Recently, a new type of symmetry for three-qubit quantum states
was introduced, the so-called Greenberger-Horne-Zeilinger (GHZ) symmetry. 
It includes the operations which leave the three-qubit  standard GHZ state unchanged.
This symmetry is powerful as it yields families of mixed states that are,
on the one hand, complex enough from the physics point of view and, on the other 
hand, simple enough mathematically so that their properties can be characterized 
analytically. We show that by using the properties of GHZ-symmetric states 
it is straightforward to derive optimal witnesses for
three-qubit entanglement.
\end{abstract}            

\vspace{2pc}

\maketitle

\section{Introduction}

Entanglement is an important resource in quantum information theory.
Therefore, the possibility to detect whether or not a given 
state is entangled is of  significant importance.
For multipartite systems, there are inequivalent
types of entanglement. A given task generally requires a certain type (or class)
of
 entanglement, and therefore also class-specific detection of
entanglement is important.

A common scheme to classify multi-party quantum states is to consider their equivalence
with respect to stochastic local operations and classical communication 
(SLOCC)~\cite{Duer2000,Bennett2001}.
The simplest multipartite case with several entanglement classes is a system
consisting of three qubits for which  
   D\"ur {\em et al.} have identified six
different SLOCC classes of pure states, and especially two different
classes of ``true'' tripartite entanglement, the Green\-berger-Horne-Zei\-linger 
(GHZ) class and the $W$
class. The representative states of these classes are the standard GHZ state
\begin{equation}
\label{eq:GHZ}
   \ket{\mathrm{GHZ}} = \oost(\ket{000}+\ket{111})
\end{equation}
and the $W$ state
\begin{equation}
\label{eq:W}
   \ket{W} \ =\  \frac{1}{\sqrt{3}}(\ket{001}+\ket{010}+\ket{100})
\end{equation}
which both are invariant under qubit permutations.

This classification can be extended to mixed states~\cite{Acin2001}. It reveals
that there is a linear hierarchy of entanglement classes. 
In this hierarchy, the GHZ class represents the ``highest''
entanglement type, followed by the $W$ class, the biseparable states $B$, and
finally the fully separable states $\Sigma$ so that we have
$\mathrm{GHZ}\supset W\supset B \supset \Sigma$.\footnote{%
         In the following we will use the names of the SLOCC classes
         in an exclusive sense, that is, we will refer to the set
         GHZ$\setminus W$ as GHZ-class states, to the set $W\setminus B$
         as $W$-class states, and to $B\setminus\Sigma$ as
         the biseparable states.} 
Mixing states of a higher class can result in a state of a lower class, but
not vice versa. 
An obvious example is that of equally mixing 
the elements of an orthogonal basis of GHZ states (GHZ
type), which gives the completely mixed state (fully separable). 
Note that, by definition, each set
of states which has \emph{at most} a given entanglement type is always
convex. 

The standard method to detect the SLOCC class of a given state are entanglement 
witnesses~\cite{Acin2001,Horodecki1996,Terhal2000b,Lewenstein2000}, 
see also~\cite{Guehne2009}  and references therein.
Despite the impressive progress in entanglement theory during the past
decade~\cite{Plenio2007,Amico2008,Horodecki2009}, to date there is still no method 
which could determine the SLOCC class of an arbitrary three-qubit state 
with certainty.  Therefore, exact results for special families of states 
are important towards a better insight into the structure
of the space of mixed quantum states. One way of obtaining such results is
the reduction of the number of free parameters by introducing certain
symmetry requirements~\cite{Werner1989,Horodecki1999,Terhal2000,Vollbrecht2002}.
Recently, a complete characterization of the entanglement type in
the family of three-qubit GHZ-symmetric states has
been achieved~\cite{Eltschka2012}. 
That family contains in particular the important case of 
pseudo-pure GHZ states, that is, mixtures
\begin{equation}
    \rho^{\mathrm{W}}\ =\ p\ \ket{\mathrm{GHZ}}\!\bra{\mathrm{GHZ}}
                        \ +\ (1-p)\ \frac{1}{8}\id
\label{eq:pseudo-pure}
\end{equation}
of the GHZ state (\ref{eq:GHZ}) 
and the state $\rho_{\mathrm{mixed}}=\frac{1}{8} \id$, moreover
$0\leq p\leq 1$.  The state $\rho_{\mathrm{mixed}}$ 
is fully unpolarized and is often regarded as a model for white noise. 

However, in
experiments one cannot always assume white noise, and for more general
admixtures the resulting state will not be GHZ symmetric.
Therefore, methods are required to determine the entanglement 
class of such states. A possible strategy is to select certain subsets
of mixed states which lead to a more adequate representation of a given type of noise 
and allows for a more precise determination of an experiment-specific
optimal witness.  This is the approach that we pursue in this article.
To this end, we first review the properties of GHZ-symmetric three-qubit
states. Subsequently, we explain the usefulness of witnesses 
whose optimality is restricted to certain subsets~\cite{Hulpke2004}. Finally we 
explicitly demonstrate how GHZ-symmetric states may serve to find 
better witnesses for pseudo-pure states, in particular for
the mixtures in equation~(\ref{eq:pseudo-pure}).
%
%

\section{GHZ-symmetric three-qubit states}
%
In this section, we summarize the properties of 
GHZ-symmetric three-qubit states~\cite{Eltschka2012}. 
In addition, we determine the set of
GHZ-symmetric states with positive partial transpose.

Density matrices with GHZ symmetry are invariant
under unitary transformations, denoted {\em GHZ symmetry operations}, which
leave the standard GHZ state equation~(\ref{eq:GHZ}) unchanged: \\
{\em (i)} arbitrary qubit permutations, \\
{\em (ii)} simultaneous three-qubit flips 
      (i.e., application of  $\sigma_x\otimes\sigma_x\otimes \sigma_x$), \\
{\em (iii)} correlated qubit rotations about the $z$ axis of the form
  \begin{equation}
    \label{eq:zrot}
    U(\phi_1,\phi_2) = \rme^{\rmi \phi_1 \sigma_z}\otimes\rme^{\rmi \phi_2 \sigma_z}\otimes\rme^{-\rmi
      (\phi_1+\phi_2) \sigma_z}\ \ .
  \end{equation}
Here, $\sigma_x$ and $\sigma_z$ are Pauli operators. Note that the operations {\em (i)--(iii)}
do not change the entanglement of an arbitrary three-qubit state.

The GHZ-symmetric three-qubit states can be parametrized as
\begin{equation}
  \label{eq:rho(x,y)}
  \rho(x, y) = \left(\frac{2y}{\sqrt{3}}
    +x\right)\pi_{\mathrm{GHZ}_+}
  + \left(\frac{2y}{\sqrt{3}}-x\right)\pi_{\mathrm{GHZ}_-}
  + \left(1-\frac{4y}{\sqrt{3}}\right)\rho_{\mathrm{mixed}}
\end{equation}
where $\pi_\psi=\ket{\psi}\!\bra{\psi}$ is the projector onto state \ket{\psi}
and we have defined the GHZ$_{\pm}$ states
\begin{equation}
\label{eq:GHZpm}
\ket{\mathrm{GHZ}_\pm}\ =\ \frac{1}{\sqrt{2}}\left(\ket{000}\pm\ket{111}\right)\ \ .
\end{equation} 
The $y$ coordinate ranges  in the interval  $-1/(4\sqrt{3})\leq y\leq \sqrt{3}/4$
while, for given $y$, the $x$ coordinate may assume values 
$|x|\leq (\sqrt{3}y/2+1/8)$.
The states thus form a plane triangle (shown in Fig.~1) with corners
$(x,y)=(-1/2,\sqrt{3}/4)$ corresponding to GHZ$_-$,
$(x,y)=(+1/2,\sqrt{3}/4)$ corresponding to GHZ$_+$, and
$(x,y)=(0,-1/(4\sqrt{3}))$ which represents the state
$\rho_r=\sum_{i=001}^{110}\ket{i}\!\bra{i}$. The completely mixed state
is located at the origin $(x,y)=(0,0)$. 
\begin{figure}[htb]
  \centering
  \includegraphics[width=.6\linewidth]{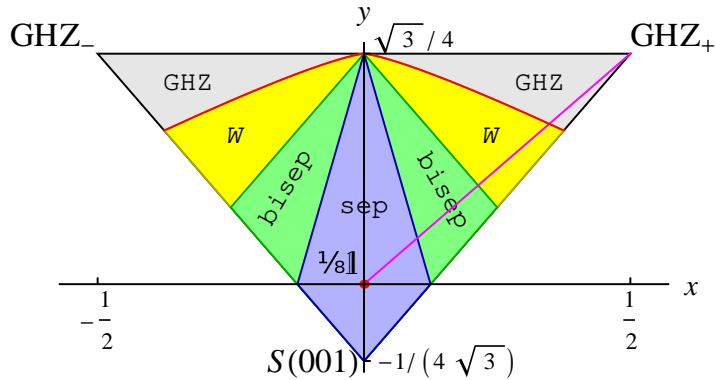}
  \caption{The set of GHZ-symmetric three-qubit states~\cite{Eltschka2012}.
           We have indicated the four SLOCC classes of three-qubit entanglement:
           GHZ (grey), $W$ (yellow), biseparable (green), and separable (blue).
           At the origin, there is the state $\rho_{\mathrm{mixed}}=\frac{1}{8} \id$.
           The solid magenta line illustrates the pseudo-pure states 
           (\ref{eq:pseudo-pure}).}
  \label{fig:ghzsymm}
\end{figure}

The GHZ-symmetric states provide an example to observe the hierarchy
of entanglement classes mentioned in the introduction.
A GHZ-symmetric state is fully separable iff it is in the polygon defined
by the four corner points $(0,-1/(4\sqrt{3}))$, $(1/8,0)$,
$(0,4/\sqrt{3})$ and $(-1/8,0)$ (the blue area in Fig.~1). Correspondingly,
it is at most 
biseparable iff it is
 in the polygon with corners $(0,-1/(4\sqrt{3}))$,
$(1/4,1/(4\sqrt{3}))$, $(0,4/\sqrt{3})$ and $(-1/4,1/(4\sqrt{3}))$ (the
green area in Fig.~1).  
A state is at most of $W$ type (yellow in Fig.~1) iff its coordinates lie in the convex 
area enclosed by 
the lower edges of the triangle and 
the curve $(x_{\mathrm{B}}(v),y_{\mathrm{B}}(v))$ given by
\begin{equation}
  \label{eq:ghzwborder}
  x_{\mathrm{B}}(v)=\frac{v^5+8v^3}{8(4-v^2)}\ \ \ ,\ \ \
  y_{\mathrm{B}}(v)=\frac{\sqrt{3}}{4}\frac{4-v^2-v^4}{4-v^2}
\end{equation}
with $-1\leq v\leq 1$. All these boundary lines
belong to the lower of the two entanglement classes which they separate
(note that the fully separable state $\frac{1}{2}(\pi_{000}+\pi_{111})$
corresponding to $(0,\sqrt{3}/4)$ is part of all three boundaries).

An arbitrary state $\rho$ can be projected onto the set of GHZ-symmetric
states by the twirling operation
\begin{equation}
  \label{eq:twirl}
  S(\rho)=\int\rmd U\, U\rho U^\dagger
\end{equation}
where the integral is understood to include also averaging over the
discrete symmetries, and $\rmd U$ is the Haar measure for the GHZ
symmetry group. The coordinates of the symmetrized state are
\begin{eqnarray}
  \label{eq:xysymm}
  x &=& \frac{1}{2}\left(\rho_{000,111} + \rho_{111,000}\right)\nonumber\\
  y &=& \frac{1}{\sqrt{3}}\left(\rho_{000,000} + \rho_{111,111} -
    \frac{1}{4}\right)\ \ .
\end{eqnarray}
Since GHZ symmetry operations do not change the entanglement class, and
mixing can result only in the same or a lower class, the entanglement class 
of $S(\rho)$ is at most that of $\rho$.

Now we determine the set of GHZ-symmetric states with positive partial
transpose (PPT). Due to the permutation invariance of the
matrix it does not matter which partition we use for the partial
transpose.
To identify the PPT states, we note that the only non-diagonal
elements of $\rho(x,y)$ are $\rho_{000,111}=\rho_{111,000}=x$. A partial
transpose obviously does not modify the diagonal elements, however it
moves those non-diagonal elements along the anti-diagonal. Since the
diagonal elements of a density matrix are always positive, the only
eigenvalues which can be negative are those of the $2\times 2$
submatrix containing $x$, that is,
\begin{equation}
  \left(
    \begin{array}{cc}
      \frac18-\frac{1}{2\sqrt{3}}y & x\\
      x & \frac18-\frac{1}{2\sqrt{3}}y
    \end{array}
  \right)
\end{equation}
The two eigenvalues of this matrix are
$\frac18-\frac{1}{2\sqrt{3}}y\pm x$, and they are both nonnegative
exactly for the polygon of fully separable states. An alternative way to obtain
this result would have been the criterion by Kay~\cite{Kay2011}.
%
%
%
%

\section{Optimal linear witnesses on subsets of states}

Linear entanglement
 witnesses are an important tool to detect 
entanglement~\cite{Horodecki1996,Terhal2000b,Lewenstein2000}, see
also~\cite{Guehne2009} and references therein.
They are defined as Hermitean operators $\mathcal{W}$ whose
expectation value $\tr(\mathcal{W}\rho)$ is positive for all unentangled
states, but negative for some entangled states. If, for some state $\rho$, 
the expectation value of $\mathcal{W}$ is negative the witness is said to detect $\rho$.

There exists a straightforward extension of this concept to
class-specific entanglement~\cite{Acin2001}: The expectation value of a class-specific
witness is positive for any state not belonging to the corresponding
entanglement class or classes, and negative for some states in those
entanglement classes. For example, a witness for GHZ-type entanglement
is negative only for states containing GHZ-type entanglement.

A useful property of linear entanglement witnesses is that they can
be specified, up to an irrelevant positive factor, solely by the set
of states for which they vanish, {\em i.e.}, a hyperplane in the space
of density matrices. This is derived directly from the corresponding
property of general Hermitean operators by noting that the expectation
value of the completely mixed state must be positive.  

To simplify terminology, in the following we will refer to states of the
entanglement type(s) which we want to detect as ``interesting'', 
whereas we will call the states
which are unentangled or have the ``wrong'' type of entanglement
``uninteresting'', respectively. In this spirit, the uninteresting states will 
always be those whose entanglement type is ''too low'' according to the
hierarchy given in the introduction, 
and therefore the set of uninteresting states will always be convex. 

An important goal is to find \emph{optimal} entanglement witnesses,
that is, witnesses which detect as many states as possible. Lewenstein
{\em et al.}~\cite{Lewenstein2000} defined a witness $\mathcal{W}$ to 
be optimal if there is no other
witness which detects all the states detected by $\mathcal{W}$ and 
also some other states not detected by $\mathcal{W}$.
Of course, this definition applies to class-specific witnesses
as well.

However, an optimal witness as defined above is not necessarily
\emph{optimal for a given set}. 
This is because a certain state may be
detected by different optimal witnesses. For example, we will see
that there are many optimal witnesses 
detecting
the standard GHZ state\footnote{%
            We will refer to these witnesses (as introduced 
            in~\cite{Lewenstein2000}) as \emph{globally optimal},
            as opposed to the witnesses which are optimal only for a
            certain subset of all three-qubit mixed states.},
but almost all of them are \emph{not} 
optimal for detecting the pseudo-pure states in equation~(\ref{eq:pseudo-pure}).
That is, there exist such mixtures which are of GHZ type, but which are not detected 
by those globally optimal witnesses. 

For this reason, we specifically study witnesses that are optimal
\emph{for a given (convex) set of states}. 
This concept has been investigated by Hulpke {\em et al.} in 
Ref.~\cite{Hulpke2004}.
To define such a witness, the
definition by Lewenstein {\em et al.}~\cite{Lewenstein2000}
 can be modified to consider only the
states in that particular set.
In the important case of mixtures of a given 
highly entangled state, such as GHZ,
with some sort of noise described by another state, the relevant set
is just a straight line in the space of density matrices 
(typically, with the entangled state in the interesting class, and the noise 
in the
 uninteresting class). In that case, this definition reduces to the
simple criterion: A witness is optimal on a given line if there is no
interesting state on that line for which the expectation value of the
witness is positive (cf.\ also Fig.~2).

Note that, using the definition above, a witness which is optimal for
a given set needs not be globally optimal:
 If the ``mixture line''
crosses a corner in the boundary of uninteresting states, for that point
there will be
generally more than one witness fulfilling the restricted optimality
condition. If, in addition, that state is also on the boundary of the
set of all states, it may happen that not all of those witnesses are globally
optimal (see Fig.~2). However, there is always a witness which is both
optimal for a given set and globally optimal.

Importantly, for any convex subset $M^{\prime}$ of the 
  set $M$ of all states, every witness $\mathcal{W}$ which is both optimal with respect to  the subset
  $M^{\prime}$ and vanishes on at least one
 full-rank state in $M^{\prime}$, 
  is also globally optimal (i.e., with respect to $M$). 
  This can be seen as
  follows: If the witness $\mathcal{W}$ were not globally optimal, there would be another
  witness $\mathcal{W'}$ that detects all the interesting states in $M$ detected by 
  $\mathcal{W}$, and then some more. Let $\rho$ be such a state with
  $\tr(\mathcal{W}\rho)>0>\tr(\mathcal{W}^{\prime}\rho)$, 
  i.e., $\rho$ is not located at the
  detection border of $\mathcal{W}$. Since $\mathcal{W}$ is optimal
  for $M'$, both witnesses must detect exactly the same states in $M'$.
  Now consider a full-rank state $\sigma\in M'$ so that $\tr(\mathcal{W}\sigma)=0$
  (which implies $\tr(\mathcal{W'}\sigma)=0$),  
  and the set of density
  matrices $\rho(\lambda)=(1-\lambda)\sigma+\lambda\rho$
  with real parameter $\lambda$. Since $\sigma$ is of
  full rank, $\rho(\lambda)$ is still a density matrix for sufficiently small
  negative $\lambda$. However for $\lambda<0$, the state is detected by
  $\mathcal{W}$ but not by $\mathcal{W'}$, in contradiction to the
  assumption.

\begin{figure}[hbt]
  \centering
  \includegraphics[width=.6\linewidth]{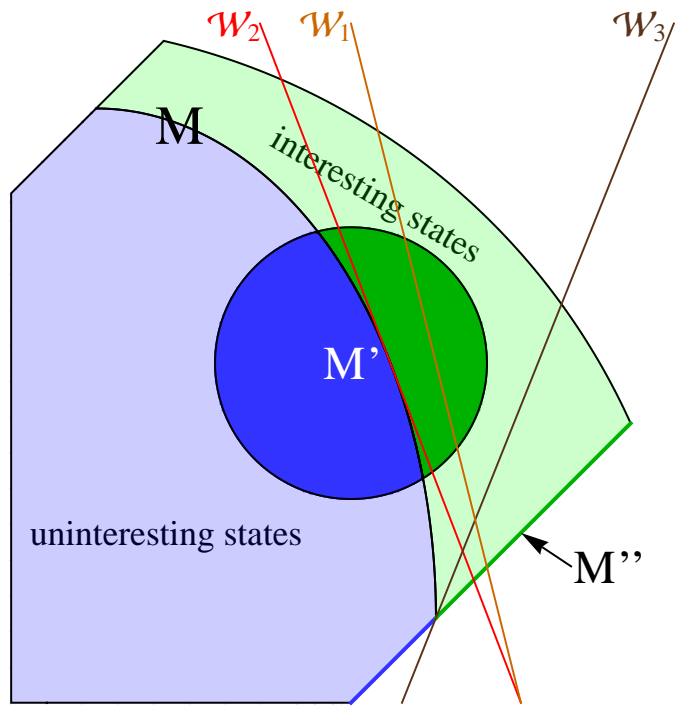}
  \caption{Illustration for different witnesses in a convex set $M$ of states which
           is a union of the disjoint sets of ``interesting'' and ``uninteresting''
           states (indicated by
           green and
 blue
 regions, respectively).
           Moreover, the circle represents a convex subset of states $M'\subset M$,
           and the right-lower border of the figure another convex subset $M''\subset M$.
           The orange line $\mathcal{W}_1$ shows a generic witness in $M$ (not optimal
           with respect to any set)
           while the red line $\mathcal{W}_2$ stands  for a witness which is both
           globally optimal in $M$ and also optimal with respect to subset $M'$
           (but not with respect to $M''$).
           The brown line  $\mathcal{W}_3$ is an example for a witness which is
           optimal with respect to the border $M''$ but
           {\em not} globally optimal (and also not optimal with respect to the circle $M'$).
           }
  \label{fig:3qbsurface}
\end{figure}
We are specifically interested in witnesses that are optimal for
the subset of GHZ-symmetric
 states. 
These states form a plane triangle that contains the completely
mixed state. 
Hence, for any witness $\mathcal{W}$ detecting a GHZ-symmetric state, 
the set
 of GHZ-symmetric states with $\tr(\mathcal{W}\rho)=0$
 is a straight line.
 As we will see, all those witnesses
 vanish on at least one full-rank GHZ-symmetric state, and are therefore
 also globally optimal.

\section{GHZ-symmetric witnesses}
We now consider a special set of witnesses, the GHZ-symmetric
witnesses, which are defined by the condition that for any GHZ
symmetry operation $U$, we have $U\mathcal{W}U^\dagger = \mathcal{W}$. It is
easy to see that such witnesses must have the form
\begin{equation}
  \label{eq:GHZsymmetricwitness}
  \mathcal{W} = a \id + b \ket{\mathrm{GHZ_+}}\!\bra{\mathrm{GHZ_+}}
  + c \ket{\mathrm{GHZ_-}}\!\bra{\mathrm{GHZ_-}}
\end{equation}
The set of states for which the GHZ-symmetric witness has zero
expectation value is then computed as
\begin{equation}
  \label{eq:witnessline}
  \tr(\mathcal{W}\rho(x, y)) = (b-c)x + \frac{\sqrt{3}}{2}(b+c)y
  + a + \frac{1}{8}(b+c) = 0 \ \ .
\end{equation}
The reverse is also true: For any straight line which does not cross
the set of uninteresting states, there exists a corresponding
GHZ-symmetric witness which is unique up to a positive factor.
\emph{Proof:} From equation (\ref{eq:witnessline}), it is immediately
obvious that for each line, there exists (up to a real factor) exactly
one GHZ-symmetric Hermitean operator 
whose expectation
 value vanishes on that line.
The condition that for the fully separable state $\rho_{\mathrm{mixed}}$
at the origin
the witness has positive expectation value fixes the sign via
the inequality $a + (b+c)/8 > 0$. 
Next we show that any such line which does not cross the set of uninteresting 
states corresponds to an entanglement witness. 
To see this, consider a state $\rho$ which is
detected by the GHZ-symmetric operator $\mathcal{W}$, {\em i.e.},
$\tr(\mathcal{W}\rho)<0$. Then the symmetrized state $S(\rho)$ is also
detected by $\mathcal{W}$, since $\tr(\mathcal{W}S(\rho)) = \int\rmd
U\,\tr(\mathcal{W}U\rho U^\dagger) = \int\rmd U\,\tr(U^\dagger\mathcal{W}U\rho) =
\int\rmd U\,\tr(\mathcal{W}\rho) = \tr(\mathcal{W}\rho)$. Now $S(\rho)$ is an
interesting state because it lies on the side of the line which is opposite to
the completely mixed state, and the line by assumption does not
cross the set of uninteresting states. As the entanglement class of
$\rho$ is at least that of $S(\rho)$, the state $\rho$ must 
also be an interesting state, which completes our proof.

From this one-to-one correspondence between 
the straight lines considered in the proof above 
and GHZ-symmetric witnesses
one also sees that for any witness
detecting GHZ-symmetric states, there is (up to a positive factor) 
a unique GHZ-symmetric witness which detects exactly the same
GHZ-symmetric states. Therefore, in order to find witnesses that are 
optimal for GHZ-symmetric states, it is justified to consider only 
GHZ-symmetric
 witnesses. From Fig.~1 it is readily seen 
that a witness is optimal for GHZ-symmetric
states iff it is tangential to the border of uninteresting states.
This is sufficient knowledge for the explicit construction 
of witnesses which are optimal for GHZ-symmetric states. 

\section{Construction of optimal GHZ-symmetric witnesses}
Since the set of GHZ-symmetric states is
symmetric under the local unitary operation $\sigma_z^{\otimes3}$ corresponding
to $\rho(x,y)\mapsto\rho(-x,y)$, it suffices to consider witnesses which detect
states with $x>0$ (since all the states $\rho(x=0,y)$ are fully
separable, there exist no witnesses detecting states with both signs
of $x$). The corresponding witness for $x<0$ can be constructed by
simply exchanging $b$ and $c$ in equation~(\ref{eq:GHZsymmetricwitness}).

We start with the detection of states which are not separable, that is, the
interesting states are those containing any type of entanglement,
while the uninteresting states are the fully separable ones. 
Here the border
is the straight line through the points $(1/8,0)$ and
$(0,\sqrt{3}/4)$, which has the equation
\begin{equation}
  \label{eq:sepborder}
  -x -\frac{1}{2\sqrt{3}}y + \frac18 = 0 \ \ .
\end{equation}
From this, together with equation~(\ref{eq:witnessline}),
the unique optimal witness (with respect to the GHZ-symmetric states)
\begin{equation}
  \label{eq:ewitness}
  \mathcal{W}_{\mathrm{bisep\setminus sep}} = \id
  - 4\ket{\mathrm{GHZ}_+}\!\bra{\mathrm{GHZ}_+}
  + 2\ket{\mathrm{GHZ}_-}\!\bra{\mathrm{GHZ}_-}
\end{equation}
is derived.

Next we consider the detection of tripartite entanglement, {\em i.e.},
the interesting states are GHZ and $W$ states, while the
uninteresting states are the biseparable and fully separable ones.
We mention that this case was studied extensively for various 
multipartite systems in Refs.~\cite{GS2010,Jungnitsch2011}.
The border between those two is again a straight line, with the
equation
\begin{equation}
  \label{eq:bisepborder}
  -2x - \sqrt{3}y + \frac34 = 0 \ \ .
\end{equation}
In analogy with the preceding case, we obtain the unique optimal witness
\begin{equation}
  \label{eq:gtewitness}
  \mathcal{W}_{W\setminus\mathrm{bisep}} = \frac{1}{2}\id - \ket{\mathrm{GHZ}_+}\!\bra{\mathrm{GHZ}_+}
\end{equation}
This witness was actually found a long time ago~\cite{Acin2001}.
It is known to be globally optimal.

Finally, we study witnesses detecting GHZ-type states. Here the
boundary is given by equation (\ref{eq:ghzwborder}), which is a convex
curve. Thus, each tangent to this curve corresponds to a witness which
is optimal for mixtures of the GHZ state and a state on the axis $x=0$
(and, since those mixtures have full rank, 
  also globally for all three-qubit states).

Due to the finite curvature of that boundary, the parameters of these witnesses 
depend on the point $(x_{\mathrm{B}},y_{\mathrm{B}})$
where the ``mixing line'' intersects the boundary.
The tangent to curve (\ref{eq:ghzwborder}) at parameter value $v=v_0$
is given by the equation
\begin{equation}
  \label{eq:ghzwtangent}
  y_{\mathrm{B}}'(v_0) x - x_{\mathrm{B}}'(v_0) y +
  \left(y_{\mathrm{B}}(v_0)x_{\mathrm{B}}'(v_0) -
    x_{\mathrm{B}}(v_0)y_{\mathrm{B}}'(v_0)\right) = 0
\ \ .
\end{equation}
From this, we obtain for the coefficients of the optimal witness
\begin{eqnarray}
  \label{eq:ghzwitnesscoefficients}
  a &=& \lambda\left(\left(y_{\mathrm{B}}(v_0) + \frac{1}{4\sqrt{3}}\right)
    x_{\mathrm{B}}'(v_0) - x_{\mathrm{B}}(v_0)y_{\mathrm{B}}'(v_0)\right)\\
  b &=& \lambda\left(\frac{1}{2}y_{\mathrm{B}}'(v_0) -
    \frac{1}{\sqrt{3}}x_{\mathrm{B}}'(v_0)\right)\\
  c &=& \lambda\left(-\frac{1}{2}y_{\mathrm{B}}'(v_0) -
    \frac{1}{\sqrt{3}}x_{\mathrm{B}}'(v_0)\right)
\end{eqnarray}
where $\lambda$ has to be chosen such that $a+(b+c)/8 > 0$. Inserting
(\ref{eq:ghzwborder}), together with an appropriate choice of $\lambda$ gives
\begin{eqnarray}
  \label{eq:ghzwitness}
  \mathcal{W}_{\mathrm{GHZ}\setminus W}(v_0) = \frac{3}{4}\id -
  \frac{3}{v_0^2-2v_0+4}\ \pi_{\mathrm{GHZ}_+} -
  \frac{3}{v_0^2+2v_0+4}\ \pi_{\mathrm{GHZ}_-}
 \ \ .
\end{eqnarray}
For the important case of the standard GHZ state mixed with white noise, 
equation~(\ref{eq:pseudo-pure}), the
mixing line crosses the boundary at $v_0=v_{\mathrm{WS}}\approx0.980701$ and thus the
optimal witness in that case reads
\begin{equation}
  \label{eq:wernerwitness}
  \mathcal{W}_{\mathrm{GHZ}\setminus W}(v_{\mathrm{WS}})\approx 0.75\ \id -
  0.999876\ \pi_{\mathrm{GHZ}_+} - 0.433327\ \pi_{\mathrm{GHZ}_-}
\end{equation}
It is instructive to compare this with the projection witness
\begin{equation}
  \label{eq:ghzpwitness}
  \mathcal{W}_{\mathrm{proj}} = \frac{3}{4} \id - \ket{\mathrm{GHZ}_+}\!\bra{\mathrm{GHZ}_+}\ \ .
\end{equation}
This witness, which happens to be GHZ diagonal as well, is derived
from the $W$-type state with maximum overlap to GHZ, which is
$\ket{W_{+--}} = (\ket{++-}+\ket{+-+}+\ket{-++})/\sqrt{3}$ (here
$\ket{\pm}=(\ket{0}\pm\ket{1})/\sqrt{2}$). The image of the symmetrized state
$\rho_{\mathrm{max}}=S(\pi_{W_{+--}})$ is located at the point $(3/8,1/({2\sqrt{3}}))$ 
in the triangle of Fig.~1 where the GHZ/$W$
 boundary reaches the lower-right border. 
This state has vanishing expectation value for the projection witness (\ref{eq:ghzpwitness}), 
therefore the projection witness is
optimal for the lower-right border of the triangle, that is, for mixtures of
GHZ with $\rho_r$. However, it is \emph{not} optimal for the entire set of
GHZ-symmetric states, as can easily be seen by calculating the optimal
witness from equation (\ref{eq:ghzwitness}). Since at the right border,
$v_0=1$, the optimal witness for all GHZ-symmetric states is
\begin{equation}
  \label{eq:optimalpmmwitness}
  \mathcal{W}_{\mathrm{GHZ}\setminus W}(1) = \frac{3}{4} \id - \ket{\mathrm{GHZ}_+}\!\bra{\mathrm{GHZ}_+}
  -\frac{3}{7}\ket{\mathrm{GHZ}_-}\!\bra{\mathrm{GHZ}_-}\ \ .
\end{equation}
This witness detects more GHZ-symmetric  GHZ-type states than the
projection witness (\ref{eq:ghzpwitness}).

\section{Conclusions}
In this article, we have described how the two-parameter family of
three-qubit GHZ-symmetric states---whose SLOCC classes of entanglement
are known exactly---can be used to derive optimal witnesses for
the SLOCC classes of general
three-qubit states. To this end, we have studied in detail 
the concept of optimal witnesses with respect to a certain subset
of states, introduced in~\cite{Hulpke2004}.  We have derived analytically 
the solutions for a bisperability witness, a witness of genuine three-qubit
entanglement and a family of optimal witnesses for GHZ-type entanglement.
Our findings confirm (witness for genuine entanglement) and improve (detection
of GHZ-type vs.\ $W$-type class) the results
for the well-known three-qubit entanglement witnesses. In particular, our approach
produces a simple one-parameter optimization scheme for witnessing GHZ-type 
entanglement in noisy GHZ states that can be applied for the assessment of
three-qubit entanglement in experimentally measured density matrices.

\ack

This work was funded by the German Research Foundation within 
SPP 1386 (C.E.), by Basque Government grant IT-472-10 and by
UPV/EHU under program UFI 11/55 (J.S.).
The authors thank O.\ G\"uhne, P.\ Hyllus and G.\ T\'oth for  helpful comments, 
and J.\ Fabian and K.\ Richter for their support.

\end{document}